\documentclass[12pt]{elsart}
\usepackage{graphicx,amssymb,amsmath}
\usepackage{listings}
\usepackage[square,comma,sort&compress]{natbib}
\usepackage[dvips,usenames]{color}

\newcommand{\functionName}[1]{{\it #1}}
\newcommand{\className}[1]{{\bf #1}}
\usepackage{fancyvrb}

\journal{Computer Physics Communications}
\begin{document}
\lstset{language=c++,basicstyle=\tiny,keywordstyle=\color{blue}\bfseries,frame=shadowbox}

\begin{frontmatter}
\title{The Density Matrix Renormalization Group for 
Strongly Correlated Electron Systems:
A Generic Implementation}
\author{G. Alvarez}
\address{Computer Science \& Mathematics  
Division and Center for Nanophase Materials Sciences, Oak Ridge National Laboratory, \mbox{Oak Ridge, TN 37831}}

\begin{abstract}
The purpose of this paper is (i) to present a generic and fully functional  implementation of the density-matrix
renormalization group (DMRG) algorithm, and
(ii) to describe how to write additional
  strongly-correlated electron models and  geometries by using templated classes.
Besides considering general models and geometries, 
the code implements Hamiltonian symmetries in a generic way and parallelization over
symmetry-related matrix blocks.
\end{abstract}
 
\end{frontmatter}

\section*{PROGRAM SUMMARY}
{\bf Manuscript title:} The Density Matrix Renormalization Group for 
Strongly Correlated Electron Systems:
A Generic Implementation\\
{\bf Author:} Gonzalo Alvarez\\
{\bf Program title:} DMRG++\\
Licensing provisions: See file LICENSE.\\
{\bf Programming language:} C++\\
{\bf Computer(s) for which the program has been designed:} PC
{\bf Operating system(s) for which the program has been designed:} Any, tested on linux
{\bf RAM required to execute with typical data:} 1GB (256MB is enough to run included test)\\
{\bf Has the code been vectorised or parallelized?:} Yes\\
{\bf Number of processors used:} 1 to 8\\
{\bf Keywords:} density-matrix renormalization group, dmrg, strongly correlated electrons, generic programming\\
{\bf PACS:} 71.10.Fd  71.27.+a  78.67.Hc\\
{\bf CPC Library Classification:} 23 Statistical Physics and Thermodynamics\\
{\bf External routines/libraries used:} BLAS and LAPACK\\
{\bf Code Location:} Computer Physics Communications\\
{\bf Development code location: }\verb!http://www.ornl.gov/~gz1/dmrgPlusPlus/!\\
{\bf Nature of problem:} 
Strongly correlated electrons systems, display a broad range of important phenomena,
and their study is a major area of research in condensed matter physics.
In this context, model Hamiltonians are used to simulate the relevant interactions of a given compound, and the relevant degrees of freedom.
These studies rely on the use of tight-binding lattice models that consider electron localization, where states on
one site can be labeled by spin and orbital degrees of freedom. 
The calculation of properties from these Hamiltonians is a computational intensive problem, since the Hilbert space
over which these Hamiltonians act grows exponentially with the number of sites on the lattice.\\
{\bf Solution method: }
The DMRG is a numerical variational technique to study quantum many body Hamiltonians.
For one-dimensional and quasi one-dimensional systems, the DMRG is able to truncate, with bounded errors and
 in a general and efficient way, the underlying Hilbert space to a constant size, making the problem tractable. \\
%
%
{\bf Running time:} The test program runs in 15 seconds.\\
{\bf References:} \cite{re:white92}\\

\section{Introduction}
In many materials of technological interest strong
interactions between the electrons lead to collective behavior.
These systems, referred to
as strongly correlated electrons systems, display a broad range of important phenomena\cite{re:alvarez07},
and their study is a major area of research in condensed matter physics.
In this context,
model
Hamiltonians are used to simulate the relevant interactions of a given compound, and the relevant degrees of freedom.
These studies rely on the use of tight-binding lattice models that consider electron localization, where states on
one site can be labeled by spin and orbital degrees of freedom. Examples of these models include the
Hubbard model\cite{re:hubbard63,re:hubbard64b}, the t-J model \cite{re:spalek77,re:spalek07}
and the spin 1/2 Heisenberg model, which can be considered the undoped limit of the t-J model. 

Non-perturbative methods to solve these fairly rich and complicated models include\cite{re:dagotto94} 
(i) quantum Monte Carlo methods, and (ii) diagonalization methods. These two paths to solve the problem are 
more or less complementary. Quantum Monte Carlo methods,
being formulated in Matsubara frequency, have difficulty obtaining real frequency properties of the model 
(such as the density-of-states), and sometimes suffer from the so-called ``sign problem"\cite{re:troyer05}. 
On the other hand,
diagonalization methods usually work efficiently only at zero or low temperature, due to the
high computational cost of obtaining a full spectrum.
Indeed, for exact diagonalization methods,
the Hilbert space over which the problem is formulated --and hence the size of the Hamiltonian matrix to be diagonalized--
 grows exponentially with the size of the system.

In 1992, S. White \cite{re:white92} introduced the density-matrix renormalization group (DMRG) method.
The DMRG is a numerical variational technique to study quantum many body Hamiltonians that could be classified as 
a diagonalization method.
For one-dimensional and quasi one-dimensional systems, this method is able to truncate, with bounded errors and
 in a general and efficient way, the underlying Hilbert space to a constant size. 
A full discussion of the DMRG is beyond the scope of the present paper, and I will only present a brief procedural description of the method.
Readers not familiar with the method are referred to the many published reviews 
\cite{re:schollwock05,re:hallberg06,re:rodriguez02}, and to the original paper \cite{re:white92}.

The present paper and accompanying code can be used in different ways. Physicists will be able to 
immediately use the flexible input file to run the
code (see Section~\ref{sec:run})
 for the Hubbard model with inhomogeneous couplings, Hubbard U values, and on-site potentials, as well as different symmetries, either on one-dimensional
chains or on n-leg ladders. 
Readers with knowledge of DMRG will be able to understand the motivation for abstraction in the implementation of the algorithm (Section~\ref{sec:generic}).
Readers with knowledge of C++ will be able to
understand the overall implementation of the DMRG algorithm (Section~\ref{sec:core}), and
write additional models (Section~\ref{subsec:models}) 
 and geometries (Section~\ref{subsec:geometries}) by following the interface provided.  Two models are included
 as examples: the Hubbard model and the spin 1/2 Heisenberg model, and
 two geometries:  the one-dimensional chain and the n-leg ladder.
Readers interested in parallelization and performance 
issues (Section~\ref{subsec:concurrency})  will be able to write other concrete concurrency classes suited to their particular hardware requirements, following
the code's parallelization abstract interface.
Finally, conclusions are presented in Section~\ref{sec:conclusions}.

Other software projects, such as the ALPS project\cite{re:albuquerque07}, also implement the DMRG algorithm
within their own frameworks.  However, this paper and DMRG++  emphasize generic programming, strongly correlated electron systems, 
detailed explanations, and few or no software dependencies.

The rest of this section is dedicated to a brief overview of the DMRG method, and to introduce some conventions and notation used throughout the paper.
Let us define {\it block} to mean a finite set of sites. 
Let $C$ denote the states of a single site. This set is model dependent. For the Hubbard model it is given by:
$C=\{e,\uparrow,\downarrow,(\uparrow,\downarrow)\}$, where $e$ is a formal element that denotes an empty state.
For the t-J model it is given by  $C=\{e,\uparrow,\downarrow\}$, and for the spin 1/2 Heisenberg model by  $C=\{\uparrow,\downarrow\}$. 
A {\it real-space-based Hilbert space} $\mathcal{V}$ on a block $B$ and  set $C$ is a 
 Hilbert space with basis $B^{C}$.  I will simply denote this as $\mathcal{V}(B)$ and assume that $C$ 
 is implicit and fixed.
A {\it real-space-based Hilbert space} can also be thought of as the external product space of $\#B$ Hilbert spaces on a site, one for each
 site in block $B$.
We will consider general Hamiltonians   that
 act on Hilbert spaces $\mathcal{V}$, as previously defined.

\begin{figure}
\centering{
\includegraphics[width=8cm]{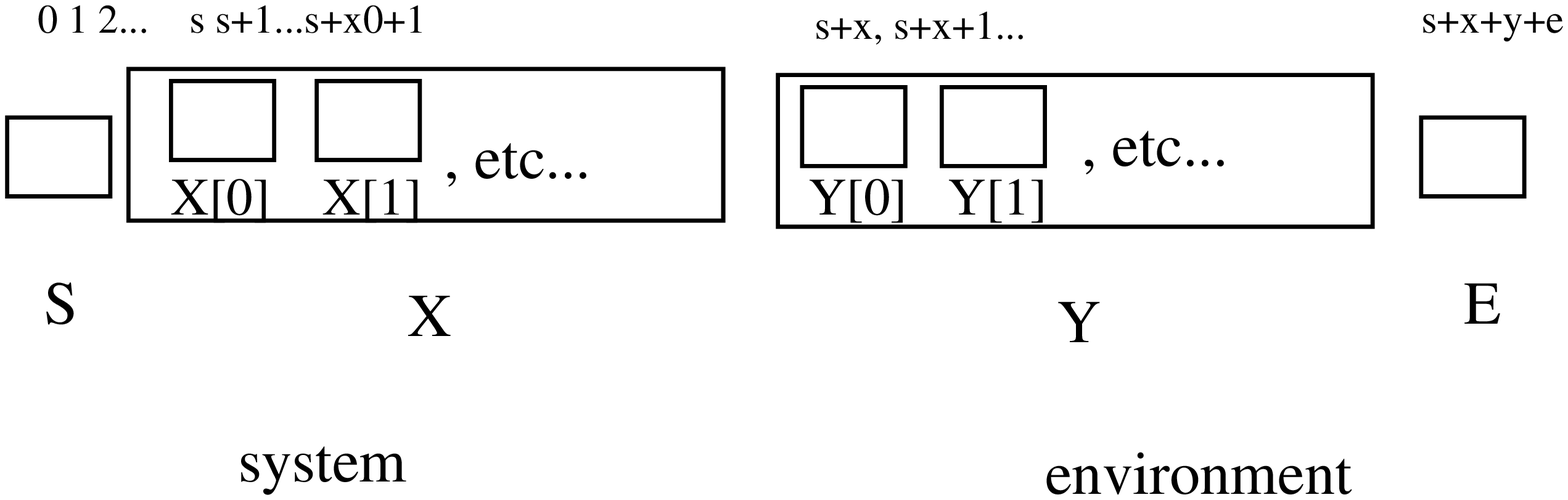}}
\caption{Labeling of blocks for the DMRG procedure. Blocks from  vector of blocks X are added one at a time to block $S$ to form 
the system and blocks from  vector of blocks
Y are added one at a time to E to form the environment. Blocks are vectors of integers. The integers (numbers at the top of the figure)
label all sites in a fixed and unique way.\label{fig:sxye}}
\end{figure}

I give a procedural description of the DMRG method in the following.
We start with an initial block $S$ (the initial system) and $E$ (the initial environment). 
Consider two sets of blocks $X$ and $Y$. 
We will be adding blocks from $X$ to $S$, one at a time, and from $Y$ to $E$, one at a time. 
Again, note that $X$ and $Y$ are sets of blocks whereas $S$ and $E$ are blocks. This is shown schematically in Fig.~\ref{fig:sxye}.
All sites in $S$, $X$, $Y$ and $E$ are numbered as shown in the figure.
Now we start a loop for the DMRG ``infinite'' algorithm
 by setting $step=0$ and $\mathcal{V}_R(S)\equiv\mathcal{V}(S)$ and $\mathcal{V}_R(E)\equiv\mathcal{V}(E)$.

The system is grown by adding the sites in $X_{step}$ to it, and let
 $S'=S\cup X_{step}$, i.e. the $step$-th block of $X$ to $S$ is added to form the block $S'$; likewise, let $E'=E\cup Y_{step}$. 
Let us form the following product Hilbert spaces:
$\mathcal{V}(S')=\mathcal{V}_R(S)\otimes \mathcal{V}(X_{step})$ and 
$\mathcal{V}(E')=\mathcal{V}_R(E)\otimes \mathcal{V}(Y_{step})$ and their union $\mathcal{V}(S')\otimes\mathcal{V}(E')$ which is disjoint.

Consider $\hat{H}_{S'\cup E'}$, the Hamiltonian operator, acting on $\mathcal{V}(S')\otimes\mathcal{V}(E')$.
We  diagonalize $\hat{H}_{S'\cup E'}$  (using Lanczos) to obtain its lowest eigenvector:
\begin{equation}
|\psi\rangle = \sum_{\alpha\in \mathcal{V}(S'), \beta\in\mathcal{V}(E')}\psi_{\alpha,\beta}|\alpha\rangle\otimes|\beta\rangle,
\label{eq:psi}
\end{equation}
where $\{|\alpha\rangle\}$ is a basis of $\mathcal{V}(S')$ and $\{|\beta\rangle\}$ is a basis of $\mathcal{V}(E')$.

Let us define the density matrices for system:
\begin{equation}
(\hat{\rho}_S)_{\alpha,\alpha'} = \sum_{\beta\in\mathcal{V}(E')}\psi_{\alpha',\beta}^*\psi_{\alpha,\beta}
\label{eq:rhoSystem}
\end{equation}
in $\mathcal{V}(S')$,
and environment:
\begin{equation}
(\hat{\rho}_E )_{\beta,\beta'}= \sum_{\alpha\in \mathcal{V}(S')}\psi_{\alpha,\beta'}^*\psi_{\alpha,\beta}
\label{eq:rhoEnviron}
\end{equation}
in $\mathcal{V}(E')$.
We then diagonalize $\hat{\rho}_S$, and obtain its eigenvalues and eigenvectors, 
$w^S_{\alpha,\alpha'}$ in $\mathcal{V}(S')$ ordered in decreasing eigenvalue order.
We change basis for the operator $H^{S'}$ (and other operators as necessary), as follows:
\begin{equation}
(H^{S' {\rm new\,\,basis}})_{\alpha,\alpha'}=(w^S)^{-1}_{\alpha,\gamma} (H^{ S'})_{\gamma,\gamma'}w^S_{\gamma',\alpha'}.
\label{eq:transformation}
\end{equation}
We proceed in the same way for the environment,  diagonalize $\hat{\rho}_E$ to obtain ordered
eigenvectors $w^E$, and define $(H^{ E' {\rm new\,\,basis}})_{\alpha,\alpha'}$.

Let $m_S$ be a fixed number that corresponds to the number of states in $\mathcal{V}(S')$ that we want to keep. 
Consider the first $m_S$ eigenvectors $w^S$, 
 and let us call the Hilbert space spanned by them, $\mathcal{V}_R(S')$, the DMRG-reduced Hilbert space on 
block $S'$. If $m_S\ge\#\mathcal{V}(S')$ then we keep all eigenvectors and there is effectively no truncation.
We truncate the matrices $(H^{S' {\rm new\,\,basis}})$ (and other operators as necessary)
such that they now act on this truncated Hilbert space, $\mathcal{V}_R(S')$.
We proceed in the same manner for the environment.%

Now we increase $step$ by 1, set $S\leftarrow S'$, $\mathcal{V}_R(S)\leftarrow\mathcal{V}_R(S')$, 
 $H_{S'}\leftarrow H_{S}$,
 and similarly for the environment, and continue with the growth phase of the algorithm.
 
In the infinite algorithm, the  number of sites in the
 system and environment grows as more steps are performed.
After this infinite algorithm, a finite algorithm is applied where the environment is shrunk at the expense of the system, and the system is grown
at the expense of the environment. During the finite algorithm phase the total number of sites remains constant allowing for a formulation
of DMRG as a variational method in a basis of matrix product states.
The advantage of the DMRG algorithm is that the truncation procedure described above keeps the error bounded and small.
Assume $m_S=m_E=m$.
At each DMRG step\cite{re:dechiara08} the truncation error $\epsilon_{tr}=\sum_{i>m} \lambda_i$, where $\lambda_i$ are the eigenvalues of the
truncated density matrix $\rho_S$ in decreasing order. The parameter $m$ should be chosen such that $\epsilon_{tr}$ remains small, say \cite{re:dechiara08}
$\epsilon_{tr}<10^{-6}$. For critical 1D systems $\epsilon_{tr}$ decays as a function of $m$ with a  power law, while for 1D system
away from criticality it decays exponentially. For a more detailed description of the error introduced by the DMRG truncation in other
systems see \cite{re:dechiara08,re:schollwock05,re:hallberg06,re:rodriguez02}.
 
\section{Motivation for Generic Programming}\label{sec:generic}
Let us motivate the discussion by introducing a typical problem to be solved by DMRG: ``Using the DMRG method, 
one would like to calculate the local density of states on all sites for a Hubbard model with
inhomogeneous Hubbard U values on a one-dimensional (1D) chain''.
We want to modularize as many tasks mentioned in the last sentence as possible. We certainly want to separate the DMRG solver from the model in question,
since we could later want to do the same calculation for the t-J model; and
the model from the lattice, since we might want to do the same calculation on, say, a n-leg ladder, instead of a 1D chain.
C++ is a computer language that is very fit for this purpose, since it allows to template classes. 
Then we can write a C++ class to implement the DMRG method (\className{DmrgSolver} class),  and template this class 
on a strongly-correlated-electron (SCE) model template, so that we can delegate all SCE model related code to the SCE model class.

However, for DmrgSolver to be able to use a given SCE model, we need a convention that such SCE model class
will have to follow.
This is known as a C++ public interface, and for a SCE model it is given in  \className{DmrgModelBase}. 
To do the calculation for a new SCE model, we simply need to
 implement all functions found in \className{DmrgModelBase}  \emph{without} changing the \className{DmrgSolver} class. 
The model will, in turn, be templated on the geometry. For example, the Hubbard model  with
inhomogeneous Hubbard U values and inhomogeneous hoppings (class \className{DmrgModelHubbard}) 
delegates all geometry related operations to a templated geometry class.
Then \className{DmrgModelHubbard} can be used for, say, one-dimensional chains and n-leg ladders \emph{without} modification.
This is done by just instantiating
 \className{DmrgModelHubbard} with the appropriate
geometry class, either \className{DmrgGeometryOneD} or 
\className{DmrgGeometryLadder}, or some other class that the reader may wish to write, which implements the interface given in
\className{DmrgGeometryBase}. 

In the following sections I will describe these different modules. Since the reader
may wish to first understand how the DMRG method is implemented, I will start with the core C++ classes that implement the method.
The user of the program
will not need to change these core classes to add functionality. Instead, new models and geometries can be added by creating implementations
for \className{DmrgModelBase} and \className{DmrgGeometryBase}, and those public interfaces will be explained next.

But for now I end this section by briefly describing the ``driver'' program  for a Hubbard model on a 1D chain.
The driver program is contained in the file main.cpp. This file is created by the configure.pl script after answering questions
related to model and geometry (see also Section \ref{sec:run}).
There, \className{DmrgSolver} is instantiated with \className{DmrgModelHubbard}, since in this case one wishes to perform the calculation for the Hubbard model.
In turn, \className{DmrgModelHubbard}  is instantiated with  \className{DmrgGeometryOneD} since now one wishes to perform the calculation on a 1D chain. 

\section{Core Classes: The DMRG Solver and Bases}\label{sec:core}
\subsection{DmrgSolver and The ``Infinite'' DMRG Algorithm}
The purpose of the \className{DmrgSolver} class is to perform  the loop for the DMRG ``infinite algorithm'' discussed before.
This class  also performs  
the ``finite algorithm''  \cite{re:schollwock05} to allow for the calculation of  
observables, such as the local density of states  of 
the cluster\footnote{In general one would want to calculate the Green function $G_{ij}(\omega)$
and this observable can be implemented in a similar way.}, defined as 
\begin{equation}
N_i(\omega)=\int_{-\infty}^{\infty}\,dt\,\langle \psi| e^{-i\hat{H}t} c^\dagger_i e^{i\hat{H}t}c_i\,e^{i\omega t}|\psi\rangle,
\end{equation}
where $|\psi\rangle$ is the ground state of the system.
The program is structured as a series of header files containing the 
implementation\footnote{Traditionally, implementation is written in cpp files that are compiled separately. However, here templates
are used heavily, and to avoid complications related to templates that some C++ compilers cannot handle,
 we choose to have only one translation unit.} with each class written in the header file of the same name, 
 and a ``driver'' program that uses the
capabilities provided by the header files to solve a specific problem.
To simplify the discussion, we start where the ``driver program'' starts, in its \functionName{int main()} function, 
which calls \functionName{dmrgSolver.main()}, whose main work is to perform  the loop for the 
``infinite'' DMRG algorithm. Let us now discuss this loop which is found in the \functionName{infiniteDmrgLoop} function, 
and is sketched in Fig.~\ref{fig:infiniteloop}.

\begin{figure}
\begin{lstlisting}
for (step=0;step<X.size();step++) {
  // grow system (a)		
  grow(pSprime,pS,X[step],model,GROW_RIGHT);
  // grow environment (b)
  grow(pEprime,pE,Y[step],model,GROW_LEFT; 
  // product of system and environment (c)   
  pSE.setToProduct(pSprime,pEprime); 				
  
  diagonalize(psi,pSprime,pEprime,pSE,
       model); // (d)			
  ns=pSprime.size();
  ne=pEprime.size();
  changeAndTruncateBasis(pS,psi,pSprime,ns,ne,
    pSE,0); // (e)
  changeAndTruncateBasis(pE,psi,pEprime,ns,ne,
    pSE,1); // (f)
    
  systemStack.push(pS); //(g)		
}
\end{lstlisting}
\caption{\label{fig:infiniteloop}Implementation of the ``infinite'' DMRG loop for a general SCE model
 on a general geometry.}
\end{figure}
In Fig.~\ref{fig:infiniteloop}(a) the system pS is grown by adding the sites contained in block X[step]. Note that X is a vector of blocks to 
be added one at a time\footnote{So X is a vector of vector of integers, and 
X[step] is a vector of integers.}.  The block X[step] (usually just a single site)
is  added \emph{to the right of} pS, hence the GROW\_RIGHT flag.  The result is stored in pSprime.
Similarly is done in Fig.~\ref{fig:infiniteloop}(b) for the environment: the block Y[step]  (usually just a single site) is added to the environment
 given in pE and stored in pEprime. 
This time the addition is
done \emph{to the left of} pE, since pE is the environment.
In Fig.~\ref{fig:infiniteloop}(c) the outer product of pSprime (the new system) and pEprime (the new environment) is made and stored in pSE. 
The actual task is delegated to the \className{DmrgBasis} class (see Section~\ref{sec:dmrgbasis}).
In Fig.~\ref{fig:infiniteloop}(d) the diagonalization of the Hamiltonian for block pSE is performed, 
and the ground state vector is computed and stored in psi, following Eq.~(\ref{eq:psi}). %
Next, in Fig.~\ref{fig:infiniteloop}(e) the bases are changed 
following Eqs.~(\ref{eq:rhoSystem},\ref{eq:rhoEnviron},\ref{eq:transformation}),  truncated if necessary, and the result is stored in 
pS for the system, and in pE, Fig.~\ref{fig:infiniteloop}(f),
for the environment. Note that this overwrites the old pS and pE, preparing these variable for the next DMRG step.

A copy of the current state of the system is pushed into a last-in-first-out stack in Fig.~\ref{fig:infiniteloop}(g), 
so that it can later be used in the finite DMRG algorithm (not discussed here, see code).
The loop continues until all blocks in vector of blocks X have been added to the initial system S, and all blocks in vector of blocks Y have been added to 
the initial environment E. We repeat again that  vector of sites are used instead of simply sites to generalize the growth process,
in case one might want to add more than one site at a time.

I will later go back to this infinite DMRG loop and discuss the implementation of
the steps mentioned in the previous paragraph (i.e., growth process,  outer products, diagonalization, change of basis and truncation).
However, some of these capabilities need first the introduction of two new C++ classes to handle operations related to Hilbert space bases.

\subsection{DmrgBasis Class: Implementation of Symmetries} \label{sec:dmrgbasis}
DMRG++ has two C++ classes that handle the concept of a basis (of a Hilbert space). The first one (\className{DmrgBasis}) handles reordering 
and symmetries in a general way, without the need to consider operators.
The second one (\className{DmrgBasisWithOperators})  does consider operators, and will be explained in
the next sub-section. The advantage of dividing  functionality in this way will become apparent later.

In any actual computer simulation the ``infinite'' DMRG loop will actually stop at a certain point.
Let us say that it stops after 50 sites have been added to the system\footnote{For simplicity, this explanatory text considers the case of 
blocks having a single site, so one site is added at a time, but a more general case can be handled by the code.}.
There will also be at this point another 50 sites that constitute the environment. 
Now, from the beginning each of these 100 sites is given a fixed number from 0 to 99. 
Therefore, sites are always labeled in a fixed way and 
their labels are always known (see Fig.~\ref{fig:sxye}).
The variable block\_ of a \className{DmrgBasis} object indicates over which sites the basis represented by this object is being built.
To explain the rest of the capability handled by the \className{DmrgBasis} class, I need to explain how symmetries are treated in the program,
and how the Hilbert space basis is partitioned.
This is explained in the following.

Symmetries will allow the solver to block the Hamiltonian matrix in blocks, using less memory, speeding up
 the computation and allowing the code to parallelize matrix blocks related by symmetry. 
 Let us assume that our particular model has $N_s$ symmetries labeled by $0\le \alpha<N_s$. 
Therefore, each element $k$ 
 of the basis has $N_s$ associated ``good'' quantum numbers 
 $\tilde{q}_{k,\alpha}$.
These quantum numbers can refer to practically anything, e.g., to number of particles with a given spin or
orbital or to the $z$ component of the spin. 
We do not need to know the details to block the matrix. However, we know that these numbers are
finite, and let $Q$ be an integer such that $\tilde{q}_{k,\alpha}<Q$ $\forall k,\alpha$. 
We can then combine all these quantum numbers into
a single one, like this: $q_k = \sum_\alpha \tilde{q}_{k,\alpha} Q^\alpha$, 
and this mapping is bijective. In essence, we combined all ``good'' quantum numbers into a single one and from now on we 
will consider that we have only one Hamiltonian symmetry called the ``effective'' symmetry, and 
only one corresponding number $q_k$, the ``effective'' quantum number.
These numbers are stored in the  member  {\it quantumNumbers} of C++ class \className{DmrgBasisImplementation}.
(Note that if one has 
100 sites or less,\footnote{This is probably a maximum for systems of correlated electrons such as the Hubbard model or the t-J model.} 
then the number $Q$ defined above is probably of the order of hundreds for usual symmetries, making this implementation very practical for
systems of correlated electrons.)

We then reorder our basis such that
its elements are given in increasing $q$ number. There will be a permutation vector associated with this reordering, that will be stored 
in the member {\it permutationVector} of class \className{DmrgBasisImplementation}. 

What remains to be done is to find a partition of the basis which labels where the quantum number changes. 
Let us say that the quantum numbers of the reordered basis states are
$\{3,3,3,3,8,8,9,9,9,15,\cdots\}$. Then we define a vector named ``partition'', such that partition[0]=0, partition[1]=4, because the quantum number 
changes in the 4th position
 (from 3 to 8), and
then partition[2]=6, because the quantum number changes again (from 8 to 9) in the 6th position, etc. 
Now we know that our Hamiltonian matrix will be composed first of a block of
4x4, then of a block of 2x2, etc.

The quantum numbers of the original (untransformed) real-space basis
 are set by the model class (to be described in Section~\ref{subsec:models}), whereas the quantum numbers of outer products are handled
by the class \className{DmrgBasis}. This can be done because if $|a\rangle$ 
has quantum number $q_a$ and $|b\rangle$ has quantum number $q_b$, then\footnote{Local symmetries must be assumed here.}
$|a\rangle\otimes|b\rangle$ has quantum number $q_a+q_b$.  \className{DmrgBasis} also knows how quantum numbers change when we change the basis: they 
do not change since the DMRG transformation 
preserves quantum numbers; and  \className{DmrgBasis} also
knows what happens to quantum numbers when we truncate the basis: quantum numbers of discarded states are discarded.
In this way, symmetries are implemented efficiently, with minimal dependencies and in a model-independent way. 

\subsection{DmrgBasisWithOperators Class and Outer Product of Operators}\label{subsec:dmrgBasisWithOperators}
C++ class \className{DmrgBasis} implements only certain functionality associated with a Hilbert space basis, as mentioned in 
the previous section. However, more capabilities related to a Hilbert space basis are needed.

C++ class \className{DmrgBasisWithOperators} inherits from DmrgBasis, and contains 
certain local operators
 for the basis in question, as well as the Hamiltonian matrix.
The operators that need to be considered here are operators necessary to compute 
the Hamiltonian across the system and environment, and to compute observables. 
Therefore, the specific operators vary from model to model.
For example, for the Hubbard model, we consider $c_{i\sigma}$ operators, that destroy an electron with spin $\sigma$ on site $i$.
For the spin 1/2 Heisenberg model, we consider operators $S^+_i$ and $S^z_i$ for each site $i$. 
In each case these operators are calculated by the model class (see Section~\ref{subsec:models}) 
on the ``natural'' basis,  and added to the basis in question with a call to 
\functionName{setOperators()}.  
These local operators are stored as sparse matrices to save memory, although the matrix type is templated and could be anything.
For details on the implementation of these operators, see \className{OperatorsBase} and the two examples provided
\className{OperatorsHubbard} and \className{OperatorsHeisenberg} for the Hubbard and Heisenberg models, respectively.
Additionally, DmrgBasisWithOperators has a number of member functions to handle operations that the DMRG method performs on 
local operators in a Hilbert space basis. 
These include functions to create an outer product of two given Hilbert spaces, to transform a basis, to truncate a basis, etc.

Let us now go back to the ``infinite'' DMRG loop and discuss in more detail Fig.~\ref{fig:infiniteloop}(a) ((b) is similar)), i.e., 
the  function grow(), which is
 found  in \className{DmrgSolver}. 
Local operators are set for the basis in question with a call to 
\className{DmrgBasisWithOperators}'s member function \functionName{setOperators()}.  
 When adding sites to the system or environment the program does a full outer product, i.e., it increases the size of 
all local operators. 
This is performed by the call to  \functionName{setToProduct(pSprime,pS,Xbasis,dir,option)} in the grow function, which actually calls
\functionName{pSprime.setToProduct(pS,xBasis,dir)}.
This function also recalculates the Hamiltonian in the outer product of (i) the previous system basis $pS$, and (ii) the basis $Xbasis$ corresponding to the
 site(s) that is (are) being added.
To do this, the Hamiltonian connection between the  two parts needs to be calculated and added, and this is done in the call to 
\functionName{addHamiltonianConnection}, found
 in the  function grow(). Finally, the resulting dmrgBasis object for the outer product, pSprime, is set to contain this full Hamiltonian with the call
to  \functionName{pSprime.setHamiltonian(matrix)}. 

I will now explain how the full outer product between two operators is implemented. 
If local operator $A$ lives in Hilbert space $\mathcal{A}$ and local operator $B$ lives in Hilbert space $\mathcal{B}$, then 
$C=AB$ lives in Hilbert space $\mathcal{C}=\mathcal{A}\otimes\mathcal{B}$.
Let $\alpha_1$ and $\alpha_2$ represent states of $\mathcal{A}$, and let $\beta_1$ and $\beta_2$
represent states of   $\mathcal{B}$. Then, in the product basis,
$C_{\alpha_1,\beta_1;\alpha_2,\beta_2}=A_{\alpha_1,\alpha_2}B_{\beta_1,\beta_2}$.
Additionally,  $\mathcal{C}$ is reordered such that each state of this outer product basis is labeled in increasing effective quantum number (see 
Section~\ref{sec:dmrgbasis}).
 In the previous example, if
the Hilbert spaces  $\mathcal{A}$ and $\mathcal{B}$ had sizes $a$ and $b$, respectively, then their outer product would have size $ab$. 
When we add sites to the system (or the environment) the memory usage remains bounded by the truncation, and it is usually not a problem to store
full product matrices, as long as we do it in a sparse way (DMRG++ uses compressed row storage). 
In short, local operators are always stored in the most recently transformed basis 
for \emph{all sites} and, if applicable, \emph{all values}  of the internal degree of freedom $\sigma$. 

This simplifies the implementation, but it must be remembered that only the 
local operators corresponding to the most recently added sites will be meaningful.
Indeed, if we  apply transformation $W$ (possibly truncating the basis, see Eq.~(\ref{eq:transformation})) then
\begin{equation}
(W^\dagger A W)  (W^\dagger BW) \neq W^\dagger  (AB)  W,  
\end{equation}
since $WW^\dagger\neq 1$ because the DMRG truncation does not assure us that $W^\dagger$ will be the right inverse of $W$ 
(but $W^\dagger W=1$ always holds). Because of this reason we cannot construct the Hamiltonian simply from
transformed local operators, even if we store them for all sites, but we need to store also the Hamiltonian in the most recently transformed 
basis\footnote{Other observables do not suffer from this problem, because they need only be computed during the finite algorithm
phase, when $WW^\dagger= 1$ holds within truncation error.}.
The fact that \className{DmrgBasisWithOperators} stores local operators in the most recently transformed basis 
  for \emph{all sites} does not increase memory usage too much, and
simplifies the writing of code for complicated geometries or connections. The SCE model class is responsible for 
determining whether a transformed operator can be used (or not because of the reason mentioned above).

Let us now examine in more detail Fig.~\ref{fig:infiniteloop}(c), where 
we form the outer product of the current system and current environment, and calculate its Hamiltonian.
We could use the same procedure as outlined in the previous paragraph, i.e., to use the DmrgBasisWithOperators class
to resize the matrices for all local operators.
Storing matrices in this case (even in a sparse way and even considering that there is truncation) would be too much of
a penalty for performance. Therefore, in this latter case we do the outer product on-the-fly only, without storing any matrices. 
In Fig.~\ref{fig:infiniteloop}(c)  pSE contains the outer product
of system and environment, but pSE is only a \className{DmrgBasis} object, not a \className{DmrgBasisWithOperators} object, i.e., it does not
contain operators. 

We now consider Fig.~\ref{fig:infiniteloop}(d), where the diagonalization of the  system's plus environment's Hamiltonian is performed.
Since pSE, being only a \className{DmrgBasis} object, does not contain all the information related to the outer product of system and environment 
(as we saw, this would be prohibitively expensive), we need to pass the system's basis (pSprime) and the environment's basis (pEprime) to
the diagonalization function (\functionName{diagonalize()} in \className{DmrgSolver}) in order to be able to form the outer product on-the-fly.
There,  since pSE does provide information about effective symmetry blocking, we block the Hamiltonian matrix using effective symmetry,
and call  \functionName{diagonaliseOneBlock()} in \className{DmrgSolver} for each symmetry block. Only those matrix blocks
that contain the desired or targeted number of electrons will be considered.
To diagonalize Hamiltonian H we use the Lanczos method\cite{re:lanczos50,re:pettifor85}, although this is also templated. 

\lstset{language=c++,basicstyle=\normalsize,keywordstyle=\color{blue}\bfseries,frame=shadowbox}
For the Lanczos diagonalization method we also want to provide as much code isolation and modularity as possible. 
The Lanczos method needs only to know how
to perform the operation $x$\lstinline! += H!$y$, given vectors $x$ and $y$. Using this fact, we can separate the matrix type from the Lanczos method.
To keep the discussion short this is not addressed here, but can be seen in the \functionName{diagonaliseOneBlock()} function, and in 
classes \className{SolverLanczos}, \className{HamiltonianInternalProduct}, and \className{DmrgModelHelper}.
The first of these classes contains an implementation of the Lanczos method that is templated on
a class that simply has to provide the operation  $x$\lstinline! += H!$y$ and, therefore, it is generic and valid for any SCE model.
It is important to remark that the operation   $x$\lstinline! += H!$y$ is finally delegated to the model in question.
As an example,  the operation \lstinline!x += H! for the Hubbard model is performed in function \functionName{matrixVectorProduct()} in class \className{DmrgModelHubbard}. 
This function performs only three tasks: (i)   $x$\lstinline! += H!$_{system}y$, (ii)  $x$\lstinline! += H!$_{environment}y$ and 
(iii)  $x$\lstinline! += H!$_{connection}y$.
The fist two are straightforward, so we focus on the last one, in \functionName{hamiltonianConnectionProduct()}, 
that considers the part of the Hamiltonian
that connects system and environment. This function runs the following loop: for every site $i$ in the system and every site $j$ in the environment
it calculates  $x$\lstinline! += H!$_{ij}y$ in function \functionName{linkProduct}, after finding the appropriate tight binding hopping value.

The function \functionName{linkProduct} is useful not only for the Hubbard model, but it is
 generic enough to use in other SCE models that include a tight binding connection of the type $c^\dagger_{i\sigma}c_{j\sigma}$,
  and, therefore, is part of a separate class,
 \className{ConnectorHopping}. Likewise, the function \functionName{linkProduct} in \className{ConnectorExchange} deals 
 with Hamiltonian connections of the type $\vec{S}_i\cdot\vec{S}_j$, and can be used by models that include that type of term, such as the 
 sample spin 1/2 Heisenberg model provided with DMRG++.
We remind readers that wish to understand this code that the function \functionName{linkProduct} and, in particular, the 
related function \functionName{fastOpProdInter} are more complicated than usual, since (i) the outer product is constructed on the fly, and (ii)
the resulting states of this outer product need to be reordered so that effective symmetry blocking can be used.
 
\section{Abstract Classes}
\subsection{The Model Interface}\label{subsec:models}
A sample SCE model, the one-band Hubbard model,
\[
\sum_{i,j,\sigma}t_{ij}c^\dagger_{i\sigma} c_{j\sigma}
+\sum_i U_i n_{i\uparrow}n_{i\downarrow} + \sum_{i,\sigma}V_i n_{i\sigma},
\]
 is implemented in class \className{DmrgModelHubbard}. 
A sample \className{DmrgModelHeisenberg} is also included for the spin 1/2 Heisenberg model $\sum_{ij}J_{ij}\vec{S}_i\cdot\vec{S}_j$.
These models 
 inherit from the 
abstract class \className{DmrgModelBase}. 
To implement other SCE models one has to implement the functions prototyped in that abstract class.
The interface (functions in \className{DmrgModelBase}) are documented in place; here I briefly describe some of them.
The \functionName{matrixVectorProduct} function needs to implement the operation  $x$\lstinline! += H!$y$. The function \functionName{addHamiltonianConnection}
 implements
the Hamiltonian connection (e.g. tight-binding links in the case of the Hubbard Model
or products $S_i\cdot S_j$ in the case of the spin 1/2 Heisenberg model) between two basis, $basis2$ and $basis3$, in the order of the outer product,
$basis1={\rm SymmetryOrdering}(basis2\otimes basis3)$. This was explained before in Section~\ref{subsec:dmrgBasisWithOperators}, 
and the examples shown by \className{DmrgModelHubbard} and \className{DmrgModelHeisenberg}
will be helpful in the implementation of other models. 
Function \functionName{setNaturalBasis} sets certain aspects of the ``natural basis'' (usually the real-space basis) on a given block.
The operator matrices (e.g., $c^\dagger_{i\sigma}$ for the Hubbard model or $S_i^+$ and $S_i^z$ for
the spin 1/2 Heisenberg model) need to be set there, as well as the Hamiltonian and the effective quantum number for
each state of this natural basis. To implement the algorithm for a fixed density, the number of electrons for each state is also needed .

\subsection{The Geometry Interface} \label{subsec:geometries}
I present two sample geometries, one for 1D chains and one for n-leg ladders in classes \className{DmrgGeometryOneD}
 and \className{DmrgGeometryLadder}.
Both derive from the abstract class \className{DmrgGeometryBase}.
To implement new geometries a new class needs to be derived from this base class,
and the functions in the base class (the interface) needs to be implemented. 
As in the case of \className{DmrgModelBase}, the interface is documented in the code, but here I briefly describe the most important functions.

The function \functionName{setBlocksOfSites}
 needs to set the initial block for system and environment, and for the vector of blocks $X$ and $Y$ to be added
to system and environment, respectively, according to the convention given in Fig.~\ref{fig:sxye}.
There are two \functionName{calcConnectorType} functions. Both calculate the type of connection between two sites $i$ and $j$,
which can be SystemSystem, SystemEnviron, 
EnvironSystem or EnvironEnviron, where the names are self-explanatory. The function \functionName{calcConnectorValue}
 determines the value of the connector (e.g., tight-binding hopping for the Hubbard model or $J_{ij}$ for the
 case of the spin 1/2 Heisenberg model) between two sites, delegating 
the work to the model class if necessary.  The function \functionName{findExtremes} determines the 
extremes sites of a given block of sites and 
the function \functionName{findReflection} finds the ``reflection'' in the environment block of a given site in the system block or vice-versa.

\subsection{The Concurrency Interface: Code Parallelization} \label{subsec:concurrency}
The \className{Concurrency} class encapsulates parallelization. Two concrete classes that implement this interface
are included in the present code. One is for serial code (\className{ConcurrencySerial} class) that does no parallelization at all, and the other one
(\className{ConcurrencyMpi} class)  is for 
parallelization based on the MPI\footnote{See, for example, http://www-unix.mcs.anl.gov/mpi/}. 
Other parallelization implementations, e.g. using pthreads, can be similarly written by implementing this interface.
The interface is described in place in class \className{Concurrency}. 
Here, I briefly mention its most important functions. Function \functionName{rank()} returns the rank of the current processor or thread.
\functionName{nprocs()} returns the total number of processors. 
Functions \functionName{loopCreate()} and \functionName{loop()} handle a parallelization of a standard loop.
Function \functionName{gather()} gathers data from each processor into the root processor.

\section{Conclusions}\label{sec:conclusions}
\begin{figure}
\centering{
\includegraphics[width=6cm]{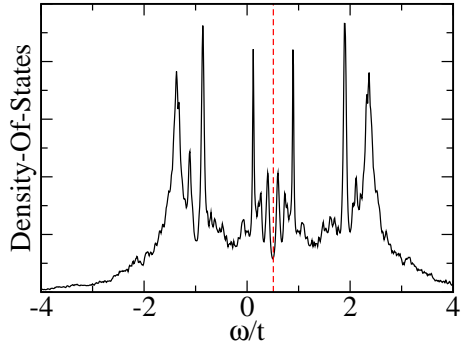}
}
\caption{Ground-State Density-Of-States of the Hubbard Model on a 32 site chain with a constant Hubbard $U=1.0$ and density 1.0.\label{fig:dos}}
\end{figure}
This paper presents DMRG++, a code to calculate 
 properties of strongly correlated electron models with the DMRG method. 
 The paper explains how to use the code for the Hubbard and spin 1/2 Heisenberg models on a one dimensional chain and on n-leg ladders, 
and how to add new models and geometries
through the use of public interfaces.
The rationale behind the design of the generic DMRG algorithm is also explained, as well as the implications for memory usage and
performance. 
The ideas used in the code --and explained in the paper-- regarding
symmetry blocking, treatment of Hamiltonian connections and parallelization, can be of inspiration to other researchers.
The code implements two efficiency techniques (suggested originally in \cite{re:white96}): (i) the wave function transformation which transforms the wave function from the previous step to
the current step to use as the initial vector for the Lanczos solver, and (ii) the use of different truncation values ``m" for different finite size loops.
Other efficiency improvements will be added to the present code in the future, for example, the use of ``disk stacks" instead of memory stacks 
(std::stack will be replaced by a DiskStack class), and the
implementation of the reflection symmetry for the infinite size algorithm (implying a factor of 2 gain during the infinite size algorithm phase).
Future work will also include a systematic treatment of ``correlation'' observables of the type $\langle \hat{O}_i\hat{O}_j\rangle$,
 which can be addressed in a generic
way.

\section{Acknowledgments}
The present code uses part of the psimag toolkit, http://psimag.org/.
Thomas Schulthess and Michael Summers's work on psimag has inspired some of the C++ templated classes used in DMRG++.
I would like to thank Jose Riera and Ivan Gonzalez 
for helping me with the validation of results and extensive tests for the DMRG code on chains and ladders.
I acknowledge the support of the 
Center for Nanophase Materials Sciences, sponsored by the Scientific User Facilities Division, Basic Energy Sciences, U.S.
Department of Energy,  
under contract with UT-Battelle.
\bibliographystyle{cpc}
\bibliography{thesis}

\begin{thebibliography}{10}

\bibitem{re:white92}
White, S.,
\newblock Phys. Rev. Lett. {\bf 69} (1992) 2863.

\bibitem{re:alvarez07}
Alvarez, G. et~al.,
\newblock J. Phys.: Condens. Matter {\bf 19} (2007) 125213.

\bibitem{re:hubbard63}
Hubbard, J.,
\newblock Proc. R. Soc. London Ser. A {\bf 276} (1963) 238.

\bibitem{re:hubbard64b}
Hubbard, J.,
\newblock Proc. R. Soc. London Ser. A {\bf 281} (1964) 401.

\bibitem{re:spalek77}
Spalek, J. and Ole{\'s}, A.,
\newblock Physica B {\bf 86-88} (1977) 375.

\bibitem{re:spalek07}
Spalek, J.,
\newblock Acta Physica Polonica A {\bf 111} (2007) 409.

\bibitem{re:dagotto94}
Dagotto, E.,
\newblock Review of Modern Physics {\bf 66} (1994) 763.

\bibitem{re:troyer05}
Troyer, M. and Wiese, U.-J.,
\newblock Phys. Rev. Lett. {\bf 94} (2005) 170201.

\bibitem{re:schollwock05}
Schollw{\"o}ck, U.,
\newblock Rev. Mod. Phys. {\bf 77} (2005) 259.

\bibitem{re:hallberg06}
Hallberg, K.,
\newblock Adv. Phys. {\bf 55} (2006) 477.

\bibitem{re:rodriguez02}
Rodriguez-Laguna, J.,
\newblock http://arxiv.org/abs/cond-mat/0207340, Real Space Renormalization
  Group Techniques and Applications, 2002.

\bibitem{re:albuquerque07}
{F. Albuquerque et al. (ALPS collaboration)},
\newblock Journal of Magnetism and Magnetic Materials {\bf 310} (2007) 1187.

\bibitem{re:dechiara08}
Chiara, G.~D., Rizzi, M., Rossini, D., and Montangero, S.,
\newblock J. Comput. Theor. Nanosci. {\bf 5} (2008) 1277.

\bibitem{re:lanczos50}
Lanczos, C.,
\newblock J. Res. Nat. Bur. Stand. {\bf 45} (1950) 255.

\bibitem{re:pettifor85}
Pettifor, D. and Weaire, D., editors,
\newblock {\em The Recursion Method and Its Applications, Springer Series in
  Solid-State Sciences}, volume~58,
\newblock Springer Verlag, Berlin/Heidelberg, 1985.

\bibitem{re:white96}
White, S.,
\newblock Phys. Rev. Lett. {\bf 77} (1996) 3633.

\bibitem{re:hallberg95}
Hallberg, K.,
\newblock Phys. Rev. B {\bf 52} (1995) 9827.

\bibitem{re:kuhner99}
{K\"uhner}, T. and White, S.,
\newblock Phys. Rev. B {\bf 60} (1999) 335.

\end{thebibliography}

\section{Test Run}\label{sec:run}
\begin{enumerate}
\item Download source code  from here \verb!http://www.ornl.gov/~gz1/dmrgPlusPlus/! (a stable version will
be published by Computer Physics Communications)
\item Create a sample driver code (main.cpp file) for the program by executing: \\
{\tt perl configure.pl}\\
and answering the questions regarding model and geometry. Defaults values can be chosen by pressing enter.
This will also create a Makefile and sample input file.
\item Run {\tt make} to compile and link the code. The LAPACK library is required by the program. File INSTALL contains more details on compilation.
\item Optionally, edit the sample input input.inp to adjust the parameters of the run. This file is self-explanatory.
\item Run the code with {\tt ./dmrg input.inp > output.dat}
\item Note1: Progress is written to standard output. Energies and
continued-fraction data is written to the file specified in input.inp:

\begin{verbatim}
parameters.numberOfKeptStates=64
parameters.linSize=8
parameters.density=1
... 
(other echo of input omitted)
#Energy=-3.57537
#Energy=-5.62889
#Energy=-7.69483
#Energy=-9.76627
...
\end{verbatim}
\item Note2: To obtain local density of states data (such as Fig.~\ref{fig:dos}): (i) run the code with the option calculateLDOS (see file README for 
a description of the options line in the input file) and (ii) process the continued fraction data as follows:
\begin{verbatim}
perl contfraction.pl data.txt 16 -4 4 0.01 > figure.dos
\end{verbatim}
where {[}-4,4{]} is the energy interval over which the local density of states calculation is to be performed and 0.01 is the
energy step increment. The method used to compute the density of states data is the continued-fraction method \cite{re:hallberg95}. Other methods,
 such as the correction vector method\cite{re:kuhner99}, have been proposed (see \cite{re:hallberg06} for a detailed review).

\item Note3: A detailed explanation of compilation instructions and required software can be found in the file INSTALL. A detailed explanation of input and output can be found in the file README.
\end{enumerate}

\end{document}